\documentclass[pdftex,twocolumn,epjc3]{svjour3}          

\RequirePackage[T1]{fontenc}

\smartqed  

\RequirePackage{graphicx}
\usepackage{latexsym}
\usepackage{amssymb, amsmath}
\usepackage{subfig}
\RequirePackage{mathptmx}      
\RequirePackage{flushend}
\RequirePackage[numbers,sort&compress]{natbib}
\RequirePackage[colorlinks,citecolor=blue,urlcolor=blue,linkcolor=blue]{hyperref}

\newcommand{\fr}{\mathtt{R}}
\newcommand{\fl}{\mathtt{L_m}}
\newcommand{\frl}{f(\mathtt{R,L_m})}

\journalname{Eur. Phys. J. C}

\begin{document}

\title{Neutron stars in $\frl$ gravity with realistic equations of state: joint-constrains with
  GW170817, massive pulsars, and the PSR J0030+0451 mass-radius from {\it NICER} data}


\author{R. V. Lobato\thanksref{e1, addr1, addr2}
        \and
        G. A. Carvalho\thanksref{e2, addr3}
        \and
        C. A. Bertulani\thanksref{e3, addr1}
}

\thankstext{e1}{e-mail: r.vieira@uniandes.edu.co}
\thankstext{e2}{e-mail: araujogc@ita.br}
\thankstext{e3}{e-mail: carlos.bertulani@tamuc.edu}

\institute{Department of Physics and Astronomy, Texas A\&M University-Commerce, Commerce, TX 75429, USA\label{addr1}
\and Departamento de Fisica, Universidad de los Andes, Bogot\'a, Colombia\label{addr2}
\and Instituto de Pesquisa e Desenvolvimento (IP\&D), Universidade do Vale do Para\'iba, 12244-000, S\~ao Jos\'e dos Campos, SP, Brazil\label{addr3}
}

\date{Received: date / Accepted: date}

\maketitle

\begin{abstract}
In this work, we investigate neutron stars (NS) in $\frl$ theory of gravity for the case $\frl = \fr
+ \fl + \sigma\fr\fl$, where $\fr$ is the Ricci scalar and $\fl$ the Lagrangian matter density. In the
term $\sigma\fr\fl$, $\sigma$ represents the coupling between the gravitational
and particles fields. For the first time the hydrostatic equilibrium equations in the theory are solved
considering realistic equations of state and NS masses and radii obtained are subject to joint constrains
from massive pulsars, the gravitational wave event GW170817 and from the PSR J0030+0451 mass-radius from
NASA's {\it Neutron Star Interior Composition Explorer} ({\it NICER}) data. We show that in this
theory of gravity, the mass-radius results can accommodate massive pulsars, while the general theory
of relativity can hardly do it. The theory also can explain the observed NS within the
radius region constrained by the GW170817 and PSR J0030+0451 observations for masses around $1.4~M_{\odot}$.
\end{abstract}

\section{Introduction}\label{sec:int}

General Relativity (GR) is by far the most successful theory of gravitation. However, in
recent years some issues came out. Data indicate that the Universe is in accelerated
expansion~\cite{riess/1998, riess/2004, des/2016}. In length scales larger than clusters of galaxies the dynamics is
governed by a negative pressure fluid, the accelerated expansion is widely accepted as caused by
a “dark energy”~\cite{bamba/2012}. Data also indicate that the galaxies have a rotation curve
flatness~\cite{rubin/1980, rubin/1985} due an invisible matter or commonly called “dark matter”. This
unknown dark energy-matter corresponds to more than 96\% of the Universe's content. In the astrophysical
level, issues have also emerged. Massive pulsars have been observed~\cite{demorest/2010,
  antoniadis/2013, linares/2018, cromartie/2020}, being hardly explained by the traditional GR theory.

One attempt to explain the emerging issues is through modified theories of gravity. Many
strategies were developed to get a theory beyond GR, some of them based on the change of the action, the Lagrangian density, or
in the metric connection. A well-studied family is the $f(\fr)$ gravity~\cite{sotiriou/2010, defelice/2010, nojiri/2017}, a tensor theory that allows
the Lagrangian to depend on higher powers of the Ricci scalar. The simplest case is just the
function $f(\fr)$ being the Ricci scalar, yielding to GR. The $f(\fr)$ theory is capable to
explain the acceleration of the Universe without dark energy~\cite{navarro/2007, song/2007,
  cognola/2008}. On the other hand, solar system tests seems to rule out most of the $f(\fr)$
models~\cite{chiba/2003, erickcek/2006, capozziello/2007a, capozziello/2008a,
  olmo/2007}. Applications of this theory have been done to neutron stars
(NS)~\cite{astashenok/2013, yazadjiev/2014, staykov/2014, yazadjiev/2015, astashenok/2015a, astashenok/2020a, astashenok/2021}, furthermore some implications of attractors
and the Higgs potential~\cite{odintsov/2021, odintsov/2021a} were taken into account in the NS
description. For the case of the Higgs potential, the WFF1 EoS, which was excluded for static neutron
stars in the context of general relativity, provides realistic results compatible with the
GW170817. It follows the same lines of reasoning as in Ref.~\cite{nunes/2020}, where is shown that one cannot rule out some EoS
without taking in consideration effects from modified theories. Although the existence of
singularities in $f(\fr)$ gravity could forbid NS formation~\cite{kobayashi/2008},
within the Palatini formalism $f(\fr)$ gravity may present optimistic results in the solar
system~\cite{toniato/2020} and for the existence of NSs~\cite{bhatti/2019, kainulainen/2007}.

Further generalizations of $f(\fr)$ gravity were developed. The $f(\mathtt{R}, \mathtt{T})$ is a
well-known case, proposed by Harko et al.~\cite{harko/2011}, it consists of a theory where
the gravitational action is an arbitrary function of the Ricci scalar and also of the trace of the
energy-momentum tensor, $\mathtt{T}$. The theory has been widely applied to compact
stars~\cite{moraes/2016, carvalho/2017, lobato/2020, pretel/2021}, see \S2.3.11 of Ref.~\cite{olmo/2020}
for a plenty of them. Notwithstanding, in most cases, the researchers considered a simple barotropic
equation of state (EoS) describing the matter inside these objects, leading to unreliable results, as
reported in Ref.~\cite{lobato/2020}. Using a set of fundamental nuclear matter EoS based
on effective models of nuclear interactions and comparing the results with the gravitational-wave
observations, as well as with massive pulsars in a joint constrain, the authors claim that the
increment in the star mass is less than 1\%. This result is in considerable contrast with previous works that
used unrealistic EoS. It also indicates that conclusions obtained from NS studies
done in modified theories of gravity without using realistic EoS that describe correctly the NS
interior, can be misleading. Another important claim is that joint constrains from electromagnetic
and gravitational-wave observations are important to rule out some extended theories of
gravity. Besides the mass' non-enhancement by the $f(\mathtt{R}, \mathtt{T})$ gravity, the
theory also incorporates a non-conservation of the energy-momentum tensor, leading to a pathology in the
hydrostatic equilibrium equations. In this regard, a non-mini\-mal geometry-matter coupling could
solve those issues~\cite{fisher/2019}. A possible theory that considers the coupling
between geometry and matter, among others~\cite{harko/2014b}, is the $\frl$ gravity, generalized by Harko and
Lobo~\cite{harko/2010}. The theory considers a general function that depends on the Ricci scalar and
also on the matter Lagrangian density $\fl$; the dynamics can only exist in the presence of matter,
satisfying the Mach's principle~\cite{dinverno/1992}. The search for the viability of such a theory has already
started in different contexts, going from studies considering it in the
  weak-field limit~\cite{harko/2010a, harko/2012} (in the weak-limit the theory could possibly converge to a
  version of MOND~\cite{barrientos/2018}), where due the theory's arbitrary coupling, extra-force
  terms appear, i.e., deviations from GR in this regime. Applications in the strong regime limit, e.g., NS, also started to show
  up~\cite{carvalho/2020a}. Here, considering NS, we shall go further than this Ref.~\cite{carvalho/2020a}, and put a window to constrain parameters from the modified gravity
perspective using realistic stellar models and realistic hadronic equations of state (EoS). The
neutron star mass-radius obtained with these EoS are subject to a joint constrain from observed massive
pulsars, the gravitational wave events GW170817, and the PSR J0030+0451 mass-radius from NASA's {\it
  Neutron Star Interior Composition Explorer} ({\it NICER}) data.

In the next section, we will briefly present the resulting hydrostatic equilibrium equation for the
underlying $\frl$ gravity theory. In section~\ref{sec:eos}, we will present the EoS that we will be using, discuss the massive pulsar observed and the results of PSR J0030+0451 from {\it NICER} data. Our results are displayed in section~\ref{sec:results}, followed by the discussion
and conclusion of our investigation in section~\ref{sec:dis}.

\section{Hydrostatic equilibrium equation in $\frl$ gravity}\label{sec:tov}
The $\frl$ gravity is a generalization of the $f(\fr)$ type gravity
models, whose action reads~\cite{harko/2010}

\begin{equation}\label{action}
S=\int d^{4}x\sqrt{-g}\, \frl,
\end{equation}
where $\frl$ is an arbitrary function of the Ricci scalar $\fr$ and of the
matter Lagrangian density $\fl$, $g$ is the metric determinant, with $8\pi G=1=c$. When the function takes the form $\frl = \fr/2 + \fl$, it conforms with the Einstein-Hilbert action, and the
variational principle leads to the well-known Einstein's field equations $G_{\mu\nu}=T_{\mu\nu}$.

Considering the simplest case where the Lagrangian density is $\frl = \fr/2 + \fl + \sigma\fr\fl$ as considered in references~\cite{garcia/2010,
  garcia/2011}, where $\sigma$ is the coupling constant, and $\fl = -p$ (pressure), the variation of the action leads to the following field equations,
\begin{eqnarray}\label{fieldeqs}
  &&(1-2\sigma p)G_{\mu\nu}+\frac{1}{3}Rg_{\mu\nu}-\frac{\sigma p}{3}Rg_{\mu\nu}\nonumber \\
   &-& (1+\sigma
  R)\left(T_{\mu\nu}-\frac{1}{3}Tg_{\mu\nu}\right) + 2\sigma \nabla_\mu\nabla_\nu p = 0.
\end{eqnarray}

To model the structure of non-rotating stars, composed of isotropic material in static
gravitational equilibrium, we consider the spherically symmetric spacetime,
\begin{equation}\label{metric}
ds^{2}=e^{\alpha}dt^2-e^{\beta}dr^{2}-r^{2}\rm{g}_{\Omega},
\end{equation}
where $\alpha$ and $\beta$ are the metric potentials depending on $r$, and $\rm{g}_{\Omega}$ is the unit 2-sphere.

Taking the energy-momentum tensor for a perfect fluid, ${\rm diag}(e^{\alpha}\rho, e^{\beta}p, r^2p, r^2\sin^2 \theta p)$. We
obtain the following components, 00 and 11 respectively, for the field equations,
\begin{subequations}\label{hee}
  \begin{eqnarray}
&&\sigma e^{-\beta} \alpha' p' - \frac{1}{3} \, {\left(\sigma p - 1\right)} R - \frac{1}{3} \, {\left(R \sigma + 1\right)} {\left(2 \, \rho + 3 \, p\right)} \nonumber \\&-& \frac{{\left(r e^{-\beta} \beta' - e^{-\beta} + 1\right)} {\left(2 \, \sigma p - 1\right)}}{r^{2}} = 0,
\end{eqnarray}
\begin{eqnarray}
&&{\left(\beta' p' - 2 \, p''\right)} \sigma e^{-\beta} + \frac{1}{3} \, {\left(\sigma p - 1\right)} R - \frac{1}{3} \, {\left(R \sigma + 1\right)} \rho \nonumber \\&-& \frac{{\left(r e^{-\beta} \alpha' + e^{-\beta} - 1\right)} {\left(2 \, \sigma p - 1\right)}}{r^{2}} = 0,
\end{eqnarray}
\noindent with primes denoting derivatives regarding the radial coordinate $r$.

The four-divergence of the energy-momentum tensor, the conserved Noether current associated with
spacetime translation, reads as~\cite{harko/2010},
\begin{equation}
\nabla^{\mu}T_{\mu\nu}=(-pg_{\mu\nu}-T_{\mu\nu})\nabla^{\mu}\ln(\sigma R),\nonumber
\end{equation}
and its local conservation yields to
\begin{equation}\label{conserv}
  p'=-(\rho+p)\frac{\alpha'}{2}.
\end{equation}

The Ricci scalar is a degree of freedom, leading to the equation
\begin{equation}\label{trace}
(1+2\sigma p)R=-(1+\sigma R)T-6\sigma \Box p,
\end{equation}
\end{subequations}
derived from the trace of the field equations. The $\Box$ operator is defined as
\begin{equation}
\Box=-e^{-\beta}\left[\frac{d^2}{dr^2}-\frac{\beta'}{2}\frac{d}{dr}+\frac{\alpha'}{2}\frac{d}{dr}+\frac{2}{r}\frac{d}{dr}\right].
\end{equation}

The hydrostatic equilibrium equations in $\frl$ gravity are given by the system of
equations~\eqref{hee}.

Making the variable change $p'=z$, we rewrite~\eqref{trace} as
\begin{equation}
  R = -\frac{8\pi T + 6\sigma B}{1 + 8\pi\sigma T + 2\sigma p}
  \end{equation}
  with $B$ being,
     \begin{equation}
      B = - e^{-\beta}(z' - \beta'z/2 + \alpha'z/2 + 2z/r),
    \end{equation}
    and $T$, the trace of the energy-momentum tensor
    \begin{equation}
      T = \rho - 3p.
\end{equation}

With the variable change, the new system of equations to be solved become:
\begin{subequations}\label{tov-like}
  \begin{eqnarray}
    \alpha'(p+\rho) + 2z = 0,
    \end{eqnarray}
    \begin{eqnarray}
      p' - z = 0,
    \end{eqnarray}
    \begin{eqnarray}
      &\Bigg[\bigg(2 r^{2} \rho e^{\beta} + \big(2  R r^{2} \rho e^{\beta} + 3 r^{2} z
             \alpha'+ 6  p r \beta' \nonumber \\
             &+ 2  \big(2  R p r^{2} + 3  p\big) e^{\beta} - 6
        p\big) \sigma - \left({\left(R - 3  p\right)} r^{2} + 3\right) e^{\beta} \nonumber \\
        &- 3  r \beta' +
        3\bigg) e^{-\beta}\Bigg]({3r^{2}})^{-1} = 0,
    \end{eqnarray}
    \begin{eqnarray}
     & \Bigg[\bigg(r^{2} \rho e^{\beta} + (R r^{2} \rho e^{\beta} + 3 \, r^{2} z \beta' + 6
       \, p r \alpha' - 6 \, r^{2} z' \nonumber \\ &- (R p r^{2} + 6 \, p) e^{\beta} + 6 \, p) \sigma + {\left(R r^{2} + 3\right)} e^{\beta} \nonumber \\
       &- 3 \, r
       \alpha' - 3\bigg) e^{-\beta}\Bigg]({3 \, r^{2}})^{-1} = 0.
\end{eqnarray}
\end{subequations}

This system~\eqref{tov-like} give us the hydrostatic equilibrium equation in $\frl$ gravity. To solve it numerically, we need to
supply an equation of state, completely determining the stellar structure. To solve the system, we
also need the
boundaries conditions.

\section*{Boundary conditions}
The boundary conditions for $\frl$ are the same as for GR, i.e., we have $p(0) = p_c$ and $\rho(0) =
\rho_c$ at the center of the star $(r=0)$, where $p_c$ and $\rho_c$ are the central values of
the pressure and energy density, respectively. The stellar surface is the point at radial coordinate $r=R$, where
the pressure vanishes, $p(R)=0$. For the metric potentials, we use $\beta(0)=0$ and
$\alpha(0)=1$. For the new variable, $z$ we use $z(0)=0$. The total mass is contained inside the radius $R$, as measured by the gravitational field
felt by a distant observer. As the boundary condition is at $r=R$, the continuity of the metric requires
that
\begin{equation}
  M = m(R) = \int_o^R4\pi r^2\rho(r)dr,
  \end{equation}
The gravitational mass is obtained similarly to standard GR calculations, once we have the energy-momentum conservation relations and the connection conditions with the exterior Schwarzschild solution.
As stated previously, to solve the system of equations~\eqref{tov-like} and obtain the mass-radius, we need to provide the
equations of state (EoS). We will focus on the ultra-dense nuclear matter EoS and on the ones
used and constrained by the Laser
Interferometer Gravitational-Wave Observatory (LIGO) detection~\cite{ligo/2018, ligo/2019}.

\section{The equation of state, massive pulsars and the {\it NICER} data}\label{sec:eos}

In our analyses we are going to follow the same methodology used in Ref.~\cite{lobato/2020}; we use only the EoS that yields a maximum mass near $2.0~M_{\odot}$ considering GR; the
EoS leading to NS mass-radius need to be within/close to the region delimited by the LIGO-VIRGO
observation~\cite{ligo/2018, ligo/2019}. We choose a set of EoS considering pure nuclear matter and
one EoS for hybrid matter (with unconfined quarks). They are labeled according to their name in the
literature. For pure nuclear matter, we have the non-relativistic ones: APR~\cite{akmal/1998},
SLy~\cite{douchin/2001} and WFF~\cite{wiringa/1988}. For
relativistic EoS, we consider the MPA~\cite{muther/1987} EoS. Finally, for the EoS containing a hybrid matter of
nucleons and quarks, we consider the ALF~\cite{alford/2005} EoS. The full description of each of
these EoS is given in \S3 of Ref.~\cite{lobato/2020}. The set of parametrization: WFF1, APR4, SLy
and MPA1 were constrained in the analysis of the gravitational wave event GW170817. The EoS lead to a maximum mass near the $2.0~M_{\odot}$ limit, however, they cannot reach the mass of the most massive pulsars recently observed. In this sense, we will use these massive
pulsar as an upper limit for the mass and see if we can reach it with a modified theory of gravity. We are going to consider the extremely massive millisecond
pulsar recently discovered by Cromartie et al.~\cite{cromartie/2020}, namely PSR J0740+6620, with
$2.14_{-0.18}^{+0.20}~M_{\odot}$ (within 95.4\% credibility interval) and the PSR J2215+5135,
a millisecond pulsar with a mass $\approx~2.27~M_{\odot}$~\cite{linares/2018},
although the technique used to measure this source is not so precise. If these measurements are confirmed in a more precise way, this
pulsar would be one of the most massive neutron star ever detected. Besides these two, the
LIGO-VIRGO collaboration, reported a coalescence involving a $22.2-24.3~M_{\odot}$ black hole and a
compact object with $2.50-2.67~M_{\odot}$, with 90$\%$ confidence~\cite{abbott/2020}. If the compact
object is a NS, this is a surprise, since no EoS with ordinary matter could explain such a mass in GR context.

Soon after the LIGO-VIRGO detection, the radius of a NS made of pure nuclear hadronic matter with a
mass of $1.4~M_{\odot}$ was constrained to be $\overline{R}_{1.4}=12.39$ km~\cite{most/2018}.
More recent results from {\it NICER}~\cite{gendreau/2016} for PSR J0030+0451 lead to the estimates: (a) a
mass of $\approx~1.44~M_{\odot}$ and equatorial radius of $\approx~R_{\rm eq}=13.02$
km~\cite{miller/2019a}; and (b) $M\approx~1.24~M_{\odot}$ and equatorial radius of $\approx~R_{\rm
  eq}=12.71$ km~\cite{riley/2019}. These {\it NICER} results can be used to tightly constrain
 parameters of the stellar structure from the modified gravity perspective and the
properties of the matter at ultra-high densities. In our investigation, we are going to use it to constrain
the coupling parameter in the non-minimal geometry-matter theory $\frl$.

\section{Results}\label{sec:results}
\begin{figure}
    \centering
    \includegraphics[scale=0.55]{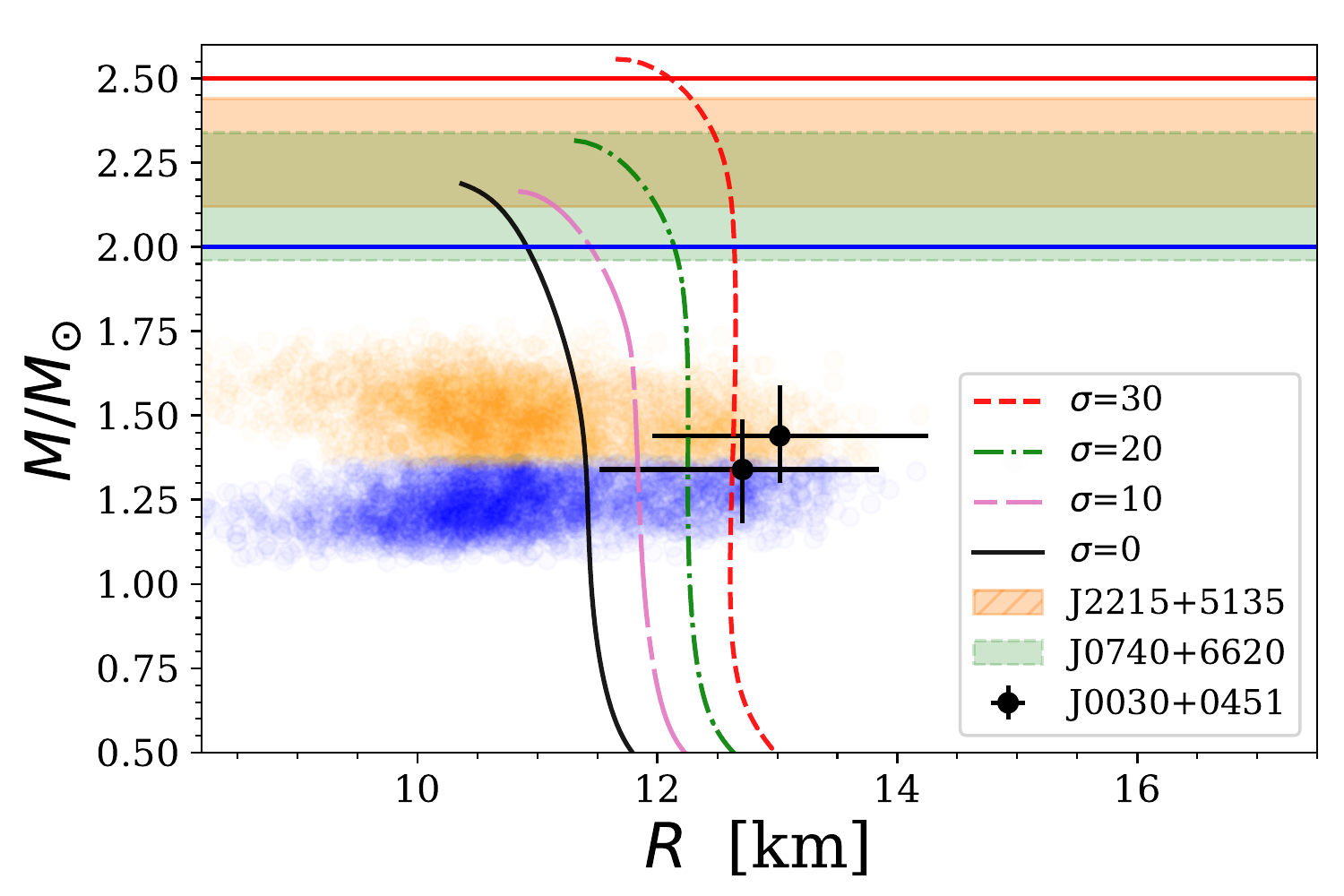}
    \caption{Mass-radius relation for the APR4 equation of state. Four values of $\sigma$ were considered.  For $\sigma=0$, the theory retrieves GR. The blue and orange clouded regions are the mass-radius constraints
   from the GW170817 event. The blue continuous line at
      2.0~$M_{\odot}$ points out the two massive NS pulsars J0348+0432 and J1614-2230, the filled
      green region represents the pulsar J0740+6620 and the filled dashed region amounts to the
      pulsar J2215+5135. The red line represents the lower mass of the compact object detected
      by the GW190414 event. The black dots with error bars, are the {\it NICER} estimations of PSR
      J0030+0451.}
    \label{fig:mr_apr4}
  \end{figure}

In figure~\ref{fig:mr_apr4} we present the mass-radius relation for the APR4 equation of state. This
EoS was constrained by experiment LIGO-VIRGO in the gravitational wave event GW170817. The
mass-radius constraints are
highlighted as the blue and orange clouded regions in the figure. The top orange region and the
bottom blue correspond to the heavier and the lighter NS, respectively. The figure also represents by a
continuous blue line the $2~M_{\odot}$ pulsars~\cite{demorest/2010, antoniadis/2013}, that we use as
a lower limit. The $2.14~M_{\odot}$ PSR J0740+6620~\cite{cromartie/2020} is shown in green filled
region as well as the $2.27~M_{\odot}$ PSR J2215+5135 in orange dashed filled region. Furthermore, we
also use the {\it NICER} mass-radius measurements~\cite{riley/2019, miller/2019a} that constrained
the mass-radius of the PSR J0030+0451, represented by blacks dots with error bars.

We generated the mass-radius curves within the $\frl$ theory of gravity for four different values of
the coupling constant $\sigma$, where $\sigma=0$ is the curve for general relativity, i.e., the $\frl$ theory
retrieves the GR. The effects of the theory are given for positive nonzero values of $\sigma$, we have used
$\sigma=10$, 20 and 30. Considering the contribution from the $\frl$ gravity, it is possible to see
an increment in the radius as the parameter increases, for $\sigma=10$ there is a small decrement in
the maximum mass, however for $\sigma=20$ and 30, there is an increment and for the later case, the
maximum mass could surpass the $2.5~M_{\odot}$ limit. Considering this EoS, the best values for $\sigma$ are $>20$,
around 30, considering the constraints, i.e., the curve is within the LIGO-VIRGO/{\it NICER} radius and can reach
the massive pulsars observed, considering the error limit. We highlight that according
to the electromagnetic counterpart of the multi-messenger observation this parametrization is
tentatively excluded~\cite{radice/2018a}.

\begin{figure}
    \centering
    \includegraphics[scale=0.55]{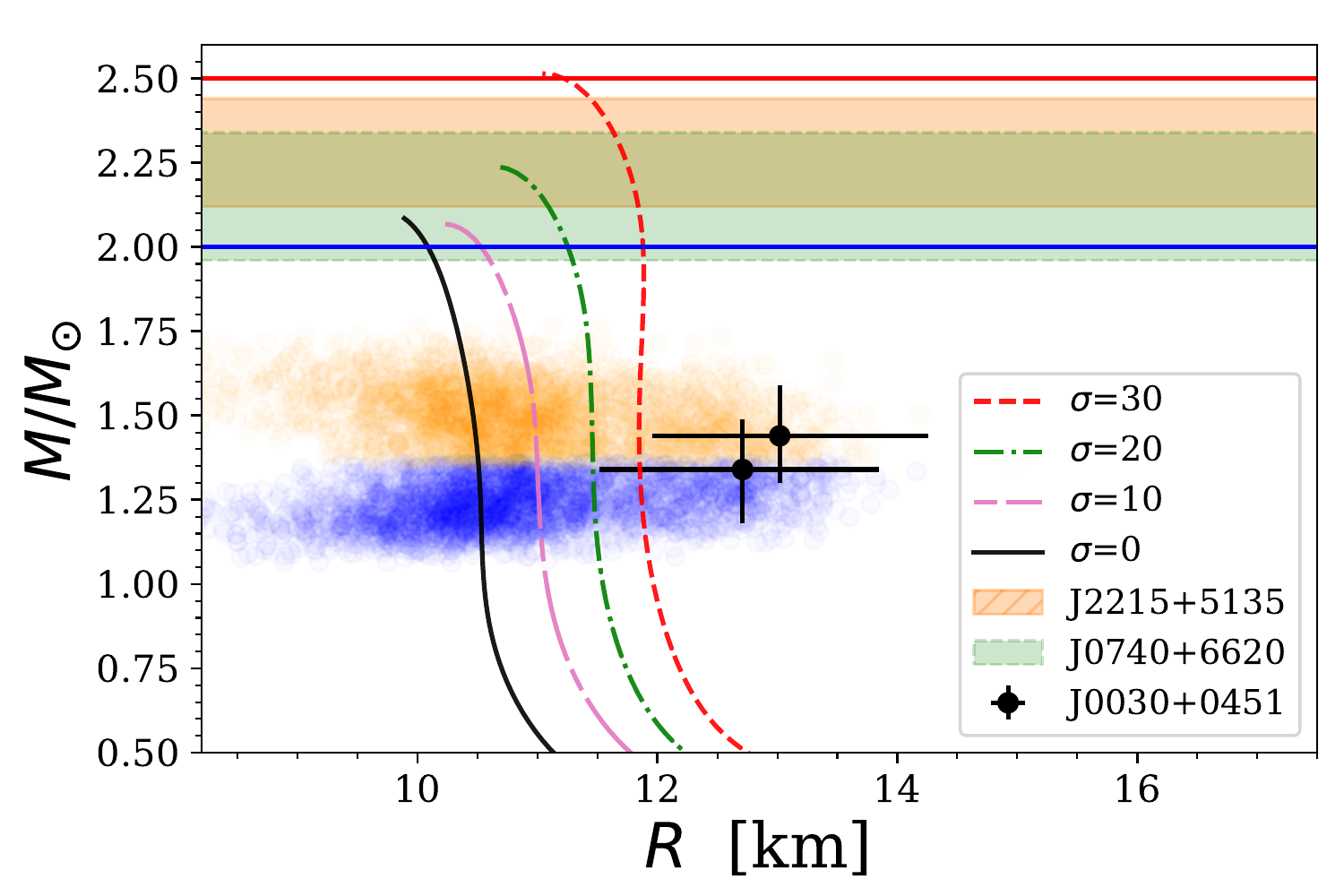}
    \caption{Same as figure~\ref{fig:mr_apr4}, but the mass-radius relation is for the WFF1 EoS.}
    \label{fig:mr_wff1}
  \end{figure}

In figure~\ref{fig:mr_wff1}, we report the mass-radius relation for the WFF1 parametrization, the
mass-radius yields a similar behavior as in the APR4 case. We observe an enhancement in the maximum mass as
we increase the parameter $\sigma$. In this EoS we obtain a small radius for the
stars, for the GR limit it is below beyond the {\it NICER} observation. The best result is achieved when the
$\sigma$ parameter is 30. However, this is still inconsistent with one of the measurements of J0030+0451. This EoS
was also disfavored by the multi-messenger observation~\cite{coughlin/2018}.

\begin{figure}
    \centering
    \includegraphics[scale=0.55]{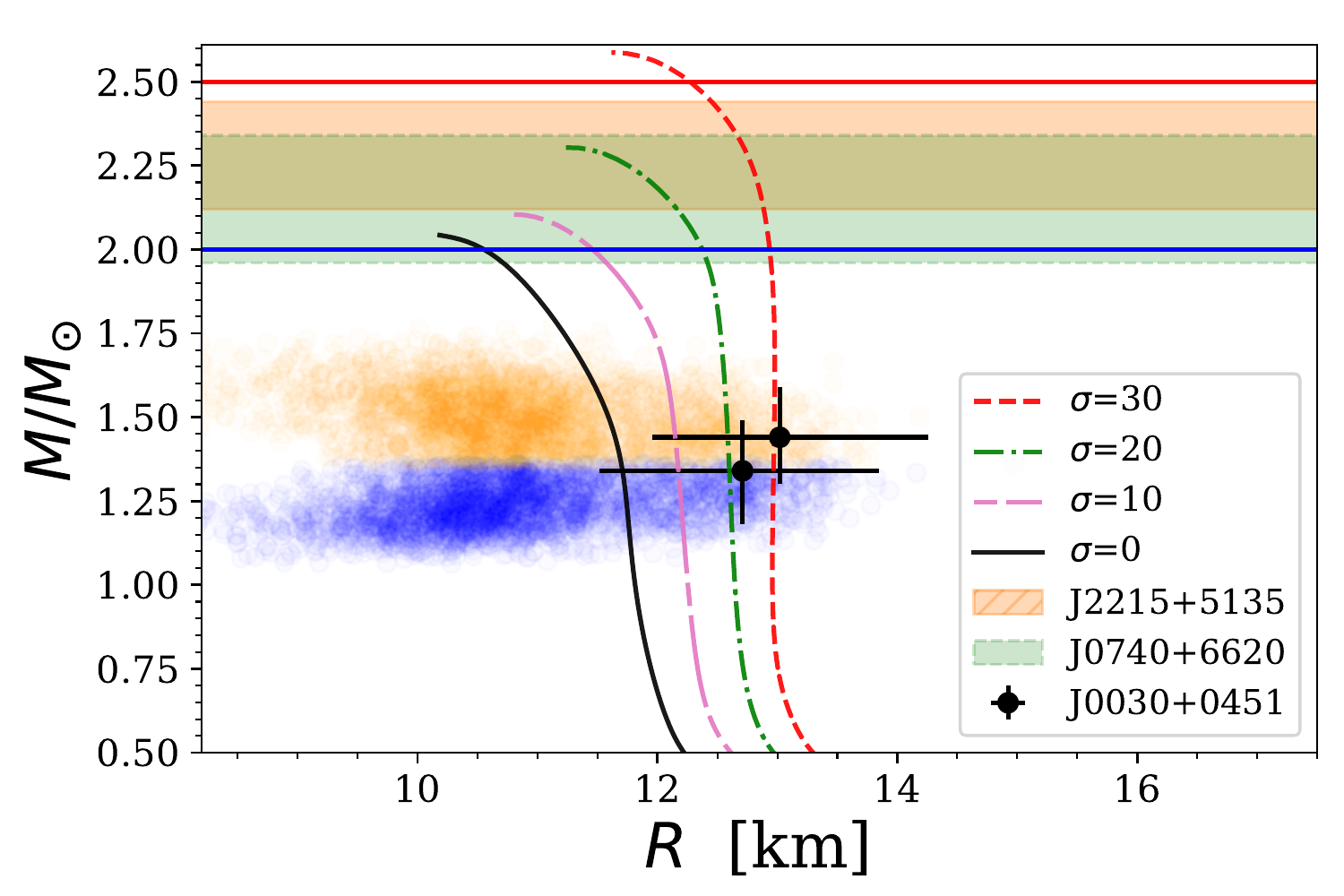}
    \caption{Same as figure~\ref{fig:mr_apr4}, but the mass-radius relation is for the SLy4 EoS.}
    \label{fig:mr_sly4}
  \end{figure}

In figure~\ref{fig:mr_sly4} we show the mass-radius relation for the SLy4 parametrization. The EoS
is largely studied in analytical representation in modified theories of gravity or GR simulations, and so on. This Skyrme type EoS can reach two solar masses when $\sigma=0$, and it is within the
LIGO-VIRGO region. But, still, it is out of rage for one of the {\it NICER} measurements. As one increases the
coupling parameter, the mass increases accordingly and for $\sigma>20$ and up to 30, the results
are well within the limits of the joint constrains, explaining the mass-radius relation for LIGO-VIRGO/{\it NICER} as
well as the mass of the massive pulsar observed.

\begin{figure}[ht]
    \centering
    \includegraphics[scale=0.55]{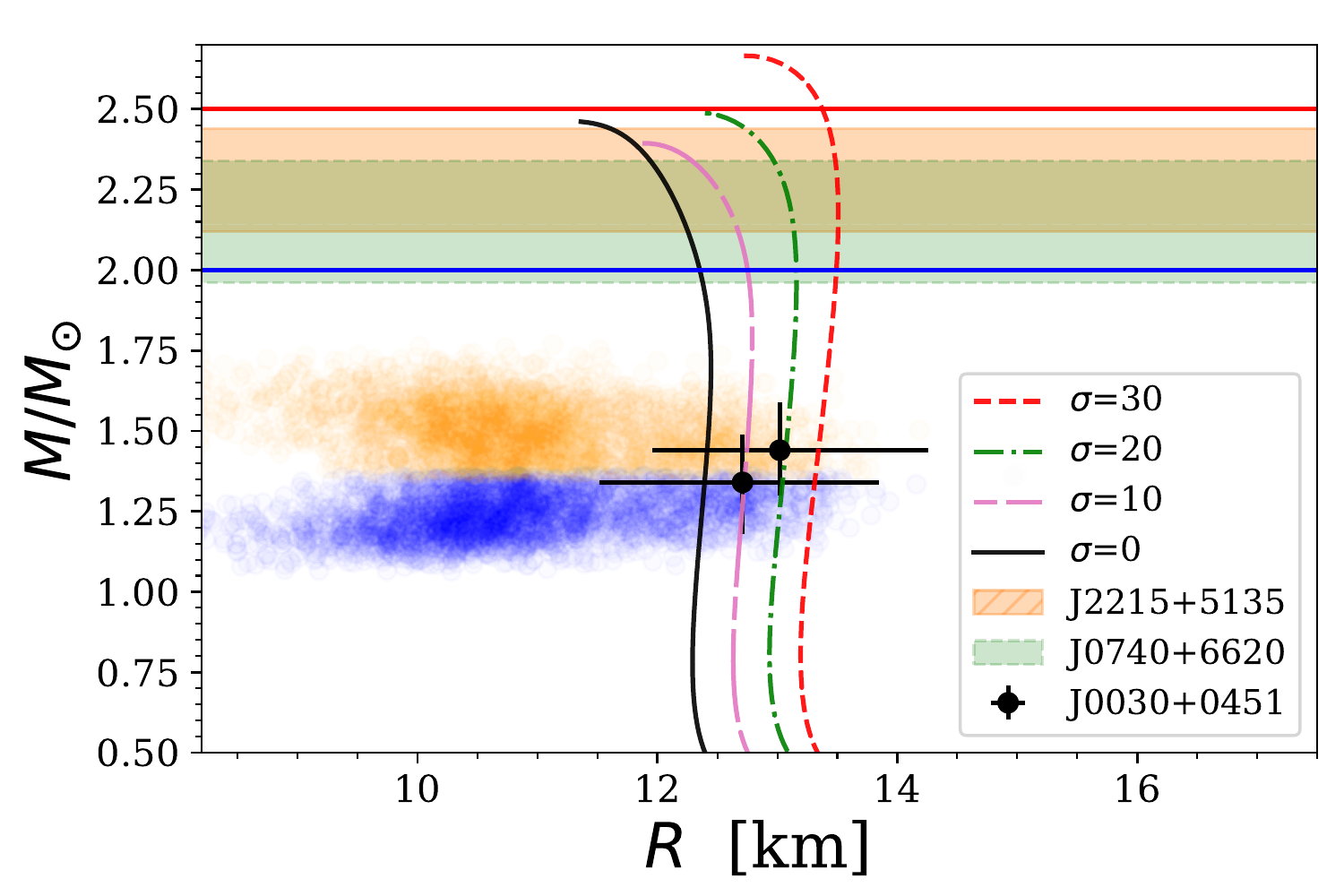}
    \caption{Same as figure~\ref{fig:mr_apr4}, but the mass-radius relation is for the MPA1 EoS.}
    \label{fig:mr_mpa}
  \end{figure}

  Figure~\ref{fig:mr_mpa} shows the mass-radius for the MPA1 parametri\-zation. This EoS can reach
  more than $2~M_{\odot}$ and is within the LIGO-VIRGO/{\it NICER} mass-radius for $\sigma=0$, i.e.,
  without any modification on the underlying theory of gravity. The curves show a similar behavior
  to APR4 and WFF1, i.e., the maximum mass point decrease at $\sigma$ around 10 and start to increase as the
  parameter enhances. The curve for $\sigma=30$ is disfavored, as it starts to lie outside the
  LIGO-VIRGO clouded region. The best value for $\sigma$ for this EoS is around 20, where the curves
  can explain all the constraints altogether. The multi-messenger observation seemed to rule out
  this EoS~\cite{ma/2018}.

\begin{figure}
    \centering
    \includegraphics[scale=0.55]{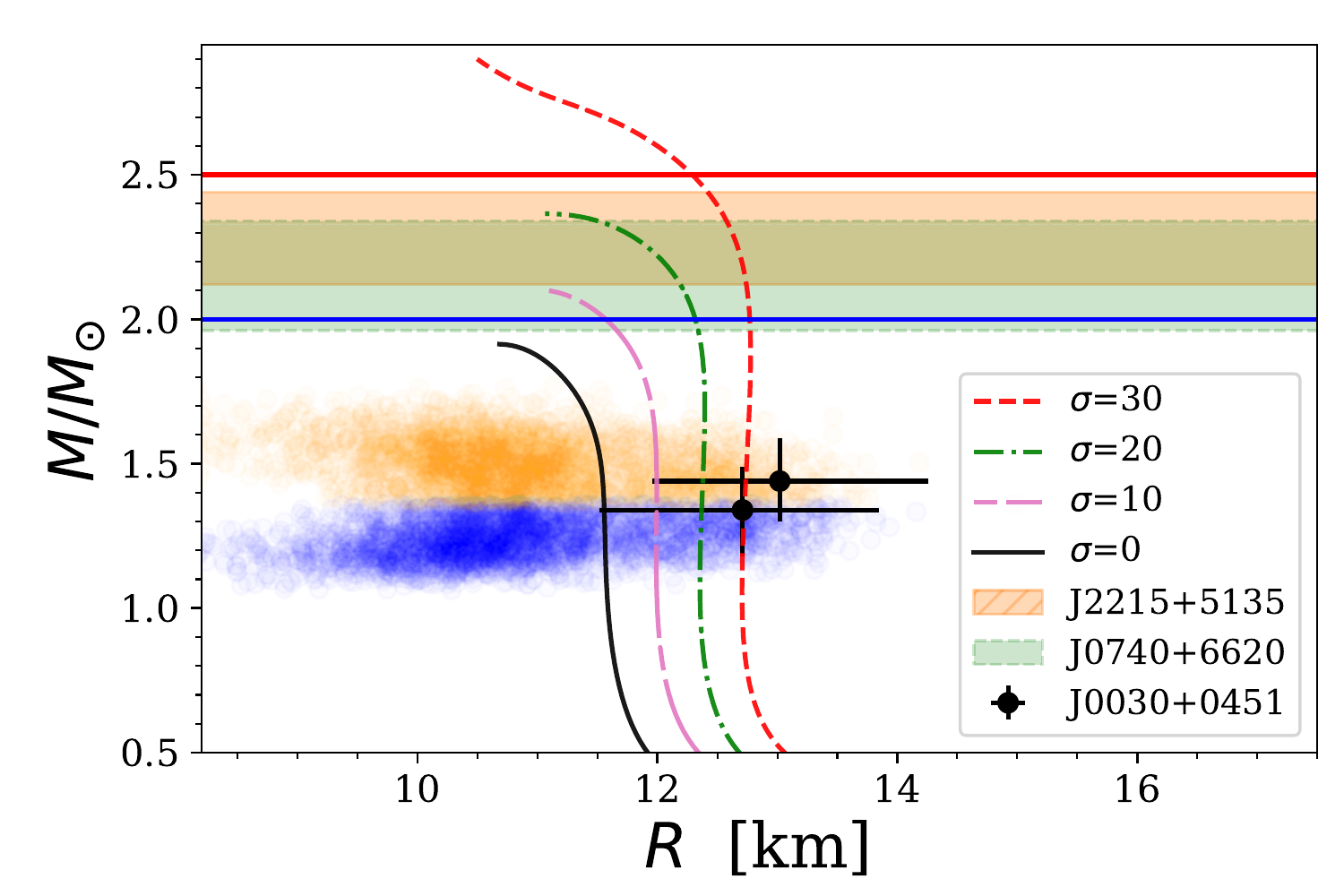}
    \caption{Same as figure~\ref{fig:mr_apr4}, but the mass-radius relation is for the ALF4 EoS.}
    \label{fig:mr_alf4}
  \end{figure}

Finally, in figure~\ref{fig:mr_alf4} the mass-radius for the ALF4 equation of state is shown. This EoS leads to the
possibility of hybrid stars. In the GR limit, the curve cannot reach the two solar mass limit and
is out of one {\it NICER} measurement. As one increases the $\sigma$ parameter, i.e., increases the effects
of the $\frl$ gravity, the maximum mass starts to increase as well. Remarkably, this EoS shows an
enhancement in the mass for different values of $\sigma$ for the same radius (see curves for $\sigma=0$ and $\sigma=30$), which is a similar behavior of
the simple barotropic equation of state~\cite{carvalho/2020a}.

  \section{Discussion and conclusion}\label{sec:dis}

In this work, we have obtained the mass-radius relationship within the non-minimal geometry-matter
coupling theory of gravity, namely the $\frl$. We have considered the simplest case:
$f=\fr+\fl+\sigma\fr\fl$, where the gravitational field is coupled to the matter field, and
$\sigma$ the coupling constant. The coupling constant presents large values in
  comparison with the weak-field limit, i.e., it is a dependence on the energy-matter density. That is
  the same dependence as in
  scalar-tensor theories, where the coupling parameter is dependent upon the background scalar field
  mass, in the so-called chameleon mechanism~\cite{khoury/2004, brax/2008, capozziello/2008a, burrage/2018}. The same mechanism appears in the non-minimal model
  $f(R,T)$, e.g., see figures in Ref.~\cite{moraes/2018c}. It would be worth to apply this theory on
  other astrophysical systems, such as black holes and white dwarfs, to study the different values of $\sigma$.

Calculations were performed for a set of EoS with
different parametrization. For the first time the hydrostatic equilibrium equations are
solved with realist equations of state considering the joint constrains from the massive pulsars
observed: (a) the gravitational wave event GW170817 from LIGO-VIRGO and (b) the mass-radius results from {\it
  NICER}. We have used EoS near to the two solar mass limit, being some of them constrained by
gravitational and electromagnetic observations. They are based on theoretical nuclear physics calculations
using many-body microscopic models fitted to numerous nuclear properties gathered in experimental data. Some EoS haven been tentatively ruled out using tidal
parameters and other gravitational wave quantities. Nevertheless, those models consider parameters adapted to the waveform coming from GR. Hence, it would be useful to
have a gravitational wave theory in $\frl$ to recalculate these wave-forms and
compare them with general relativity, and with that, maybe have new tidal
parameters and other quantities derived from gravitational waves, to apply to
neutron stars. Those parameters can change in the modified theory or simply not, so that the
  constraints we obtain here may vary. Gravitational wave emission and other study topics in strong
  regime should be
  addressed in $\frl$ gravity considering NS mergers. We can anticipate that in the vacuum,
  gravitational wave solutions will not change once the $\frl$ functional will become only
  $\mathtt{R}$ because $\mathtt{L_m}\rightarrow 0$, reducing to GR. So, the gravitational wave parameters may change as macroscopic parameters change, according to $\frl$ gravity, but gravitational wave propagation is not expected to differ from GR.

We show that the $\frl$ modified theory of gravity can account for the enhancement of the maximum mass, as
the theories' coupling constant increases. The stars' radii also increases, the increment of the
radius goes into the
inner region of the {\it NICER} results, i.e., the modified theory is in better agreement with the
observations than GR theory.

\begin{acknowledgements}
R.V.L. and C.A.B. are supported by U.S. Department of Energy (DOE) under
grant DE--FG02--08ER41533 and to the LANL Collaborative Research Program by Texas A\&M System National
Laboratory Office and Los Alamos National Laboratory. G.A.C. is supported by Coordena\c c\~ ao de Aperfei\c coamento de Pessoal de N\'ivel Superior (CAPES) grant PN\-PD/88887.368365/2019-00.
\end{acknowledgements}

\bibliographystyle{spphys}
\bibliography{lib}
\end{document}